\documentclass[letter]{aa} 

\usepackage{graphicx}
\usepackage{txfonts}
\begin{document}

   \title{
   First detection of methanol towards a post-AGB object, HD\,101584
   }

   \titlerunning{Methanol in HD\,101584}

   \author{H. Olofsson  \inst{1}
          \and
          W.H.T. Vlemmings \inst{1}
          \and 
          P. Bergman   \inst{1}
          \and      
          E.M.L. Humphreys \inst{2}
          \and
          M. Lindqvist \inst{1}
          \and
          M. Maercker \inst{1}
          \and
          L. Nyman \inst{3,4}
          \and
          S. Ramstedt \inst{5}
          \and 
          D. Tafoya  \inst{1}
          }

   \institute{Dept. of Space, Earth and Environment, Chalmers Univ. of Technology,
              Onsala Space Observatory, SE-43992 Onsala, Sweden\\
              \email{hans.olofsson@chalmers.se}
         \and
         ESO, Karl-Schwarzschild-Str. 2, D-85748 Garching bei M{\"u}nchen, Germany
         \and
         Joint ALMA Observatory, Alonso de Cordova 3107, Vitacura, Santiago de Chile, Chile
         \and
         ESO, Alonso de Cordova 3107, Vitacura, Santiago, Chile 
         \and
         Dept. of Physics and Astronomy, Uppsala University, Box 516, SE-75120 Uppsala, Sweden     
             }

   \date{Received 14 May 2017; accepted 19 June 2017}

\abstract{The circumstellar environments of objects on the asymptotic giant branch and beyond are rich in molecular species. Nevertheless, methanol has never been detected in such an object, and is therefore often taken as a clear signpost for a young stellar object. However, we report the first detection of CH$_3$OH in a post-AGB object, \object{HD\,101584}, using ALMA. Its emission, together with emissions from CO, SiO, SO, CS, and H$_2$CO, comes from two extreme velocity spots on either side of the object where a high-velocity outflow appears to interact with the surrounding medium. We have derived molecular abundances, and propose that the detected molecular species are the effect of a post-shock chemistry where circumstellar grains play a role. We further provide evidence that HD\,101584 was a low-mass, M-type AGB star.
}


   \maketitle
%
%
%
\section{Introduction}

The circumstellar envelopes (CSEs) of asymptotic giant branch (AGB) and post-AGB objects have turned out to be rich in different molecular species; more than 100 are now detected. This is the effect of a number of different processes, such as stellar atmosphere equilibrium chemistry, extended atmosphere non-equilibrium chemistry, and photo-induced circumstellar chemistry \citep[e.g.,][]{mill16}. The post-AGB objects have a special niche in terms of chemistry because of the increased internal UV light and the presence of shocks where fast winds interact with slower-moving material. The result being that they often show molecular species that are not detected (or tend to be much weaker) in AGB CSEs; for example, a number of ions \citep{sacoetal15}. One species, methanol (CH$_3$OH), has escaped every attempt at detection in an AGB-related object \citep[e.g.,][]{charlatt97, gomeetal14}, despite detectable predicted abundances \citep{willmill97}. It is therefore taken as a clear signpost for a young stellar object (YSO) \citep{breeetal13}. However, in the course of a chemical study of an interesting post-AGB object HD\,101584 using ALMA, we have detected methanol for the first time in an AGB-related object, and in this {\it Letter} we discuss its origin together with detections of CO, SiO, SO, CS, and H$_2$CO.

HD\,101584 is a bright star ($V$\,$\approx$\,7$^{\rm m}$) of spectral type A6Ia \citep{sivaetal99}. It was shown to have a large far infrared excess and an evolutionary status at, or shortly after, the end of the AGB was proposed by \citet{partpott86} and further corroborated by \citet{bakketal96a}. It has also been shown to be a binary system \citep{bakketal96b, diazetal07}. The distance is estimated to be 0.7\,kpc, but recent Gaia data suggests a somewhat larger distance, 0.9 -- 1.8\,kpc. The HST images show only a diffuse circumstellar medium in dust-scattered light \citep{sahaetal07}, but its circumstellar gas characteristics are remarkable.

\citet{olofetal15} used ALMA $^{12}$CO($J$\,=\,\mbox{2--1}) data to identify a narrow, $\approx$\,10\arcsec\ long, high-velocity molecular outflow directed at PA\,$\approx$\,90$^\circ$. Its velocity range covers almost 300\,km\,s$^{-1}$ and has a Hubble-like gradient. The outflow is seen almost along its axis. The kinematical age is estimated to be $\la$\,500\,yr. There is an hour-glass structure surrounding the outflow, and a complex structure within 1\arcsec\ (in radius) of HD\,101584, most likely a torus-like component centered on a circumbinary disc. Here we provide further evidence, based on circumstellar isotopolog ratios, for the post-AGB nature of HD\,101584.

\section{Observations}

The ALMA data were obtained during Cycles 1 and 3 with 35 to 43 antennas of the 12\,m main array in two frequency settings in Band 6, one for the $^{12}$CO($J$\,=\,\mbox{2--1}) line and one for the $^{13}$CO($J$\,=\,\mbox{2--1}) line (only Cycle~1). In both cases, the data set contains four 1.875\,GHz spectral windows with 3840 channels each. The baselines range from 13 to 12934\,m. 
\citet{olofetal15} concluded that very little flux is lost even in the $^{12}$CO(\mbox{2--1}) ALMA data, in particular at the extreme velocities and close to the systemic velocity. 
Bandpass calibration was performed on J1107-4449, and gain calibration on J1131-5818 (Cycle~1) and J1132-5606 (Cycle~3). Flux calibration was done using Ceres and Titan. Based on the calibrator fluxes, we estimate the absolute flux calibration to be accurate to within 5\%.

The data were reduced using CASA 4.5.2. After corrections for the time and frequency dependence of the system temperatures, and rapid atmospheric variations at each antenna using water vapour radiometer data, bandpass and gain calibration were done. For the $^{12}$CO($J$\,=\,\mbox{2--1}) setting, data obtained in three different configurations were combined. Subsequently, for each individual tuning, self-calibration was performed on the strong continuum. Imaging was done using the CASA clean algorithm after a continuum subtraction was performed on the emission line data. The final line images were created using natural weighting. 

Complementary $^{12}$CO $J$\,=\,\mbox{3--2} and \mbox{4--3}, and C$^{17}$O and C$^{18}$O $J$\,=\,\mbox{2--1} line data were obtained using APEX. The Swedish heterodyne facility instruments APEX-1,2 \citep{vassetal08} and APEX-3  were used together with the facility FFT spectrometer covering about 4\,GHz. The observations were made from August to October 2015 in dual-beamswitch mode with a beam throw of 2$\arcmin$. Regular pointing checks were made on strong CO line emitters and continuum sources. Typically, the pointing was found to be consistent with the pointing model within 3$\arcsec$. The antenna temperature, $T_{\mathrm A}^{\star}$, is corrected for atmospheric attenuation. A Jy/K conversion of 40 and 48 was adopted for APEX-1,2 and APEX-3, respectively. The uncertainty in the absolute intensity scale is estimated to be about $\pm 20$\%.

Finally, in our analysis we make use of the $^{12}$CO and $^{13}$CO $J$\,=\,\mbox{1--0} data published by \citet{olofnyma99}.

   \begin{figure}
   \centering
  \includegraphics[width=8cm]{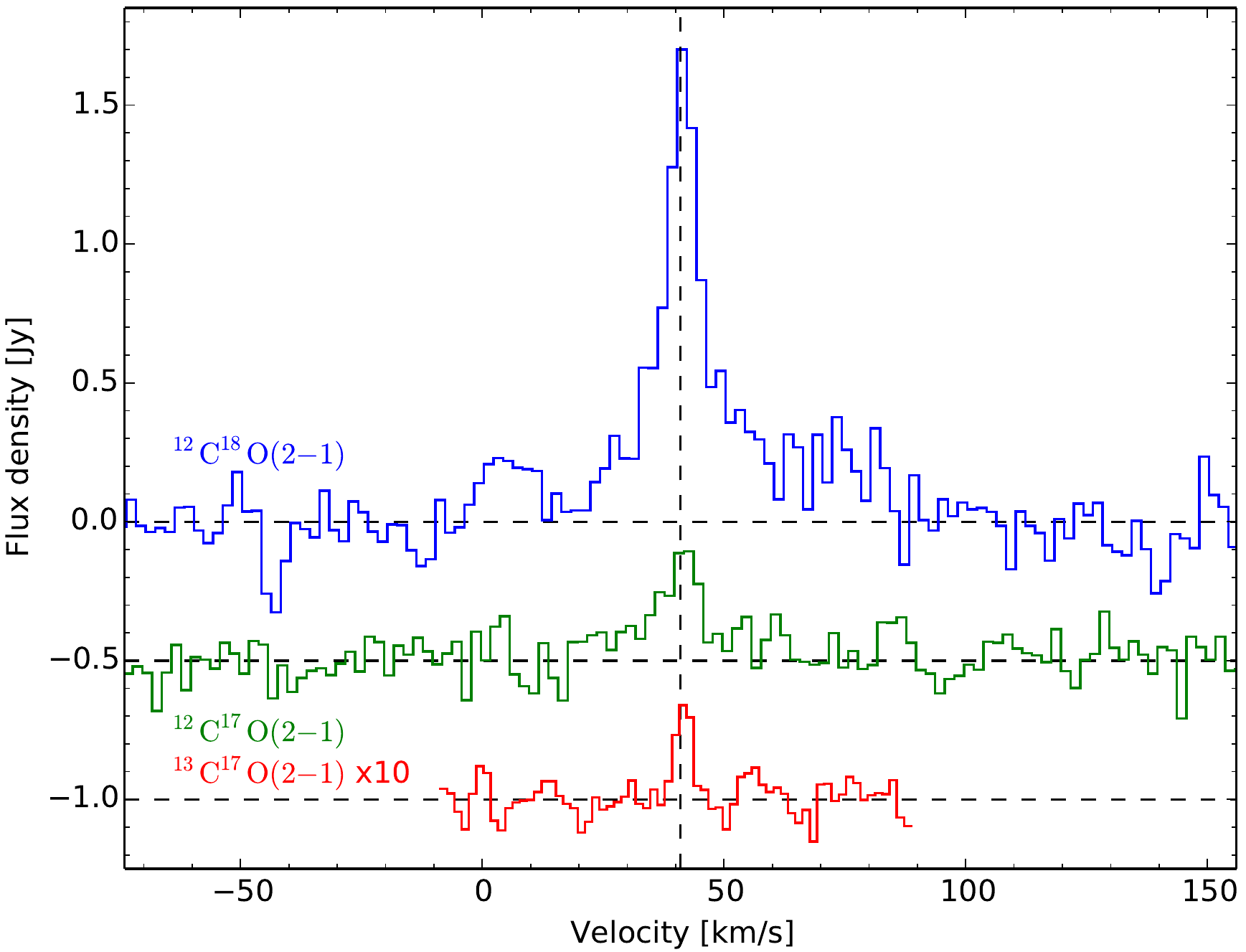}
      \caption{APEX C$^{18}$O (upper) and C$^{17}$O (middle),  and ALMA $^{13}$C$^{17}$O (lower) $J$\,=2--1 line profiles towards HD101584. The vertical dashed line marks the estimated systemic velocity. A Local Standard of Rest velocity scale is used.
              }
         \label{f:colines}
   \end{figure}   

   \begin{figure*}
   \centering
  \includegraphics[width=18cm]{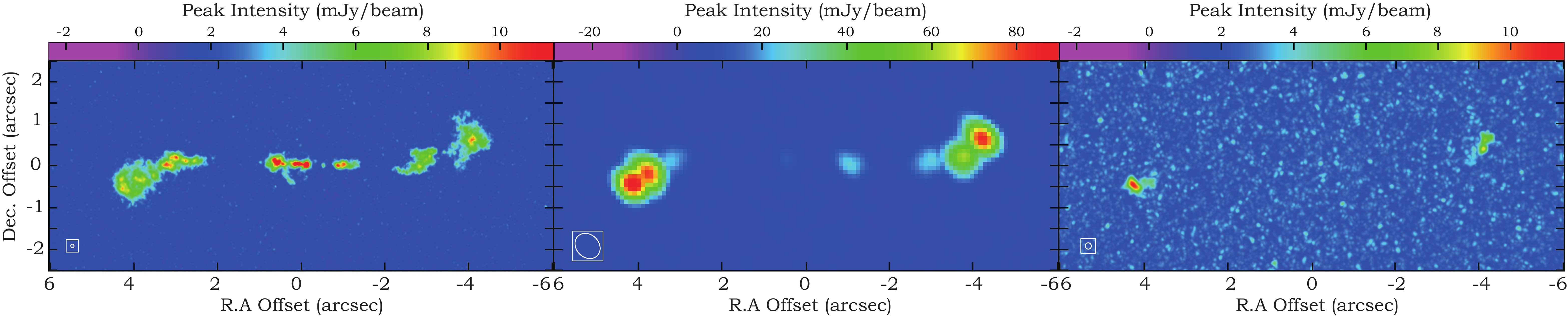} 
      \caption{SiO(5--4) (left, 0\farcs08 beam in the lower left corner), H$_2$CO(3$_{03}$--2$_{02}$) (middle, 0\farcs6 beam), and CH$_3$OH(8$_{-1}$--7$_0$) (right, 0\farcs15 beam) images towards HD101584 produced by selecting the peak intensity (in absolute terms along the velocity axis) in each pixel. This highlights the morphology of the high-velocity outflow. In particular, the methanol line emission marks the positions of the 4\arcsec\,W and 4\arcsec\,E EVSs. 
              }
         \label{f:maps}
   \end{figure*}   

\section{Discussion}

\subsection{The evolutionary nature of HD\,101584}
\label{s:nature}

The most likely interpretation of the A6Ia spectral classification and the circumstellar characteristics is that HD\,101584 is a post-AGB object,  but an evolved massive supergiant or possibly a star of young age remain alternatives. However, HD\,101584, being a high-latitude source, is not associated with any star-forming region or molecular cloud. Nevertheless, since here we claim the first detection of methanol in an AGB-related object, we provide some further evidence that supports the post-AGB nature of HD\,101584.  

The circumstellar chemistry (detections of e.g., SO, SO$_2$, and OCS, Olofsson et al. in prep.), the strong 1667\,MHz OH maser \citep{telietal92}, and the presence of a silicate feature \citep{bakketal96a} all strongly favor the idea that the circumstellar medium of HD\,101584 is O-rich (C/O\,$<$\,1). 

A stellar $^{12}$C/$^{13}$C ratio that is not consistent with that in the local interstellar medium, 45--70 \citep[e.g.,][]{lucalisz98}, nor with that of the Sun, 87 \citep{scotetal06}, would be a strong argument for an AGB-related object. We provide such an estimate where opacity and chemistry are expected to have little effect by using the APEX and ALMA detections of the $^{12}$C$^{17}$O and $^{13}$C$^{17}$O $J$\,=\,\mbox{2--1} lines, respectively (Fig.~\ref{f:colines}).  We have determined the intensities of the emission coming from the very central region, within 0\farcs5 of HD\,101584 in the ALMA data and using the narrow feature in a Gaussian line decomposition of the single-dish data (Table~\ref{t:isodata}). The integrated line intensity ratio is 13$\pm$6, suggesting a low $^{12}$C/$^{13}$C ratio in line with matter that has been processed in the CNO-cycle and brought to the surface in an AGB star;  \citet{ramsolof14}, for example, find circumstellar $^{12}$CO/$^{13}$CO ratios in the range 6--30 for M-type AGB stars, an effect believed to be due to evolution on the Red Giant Branch.  

The $^{17}$O/$^{18}$O ratio is a measure of the initial mass of an AGB star provided that it is not affected by hot-bottom-burning (HBB), that is, $M_{\rm i}$\,$\la$\,4\,M$_\odot$ \citep{hinketal16, denuetal17}. During HBB the $^{17}$O/$^{18}$O ratio rapidly becomes very high \citep{justetal15}.  The intensity ratio of the APEX C$^{17}$O and C$^{18}$O $J$\,=\,\mbox{2--1} lines (Fig.~\ref{f:colines}), 0.20$\pm$0.08, is expected to be a good measure of the $^{17}$O/$^{18}$O ratio. This shows that HD\,101584 has not gone through HBB, and the low ratio suggests an initial mass of $\approx$\,1\,M$_\odot$ \citep{denuetal17}. This low mass is also consistent with the fact that HD\,101584 has not evolved into a carbon star, and with the present-day mass estimate of \citet{bakketal96a} provided that only the core mass remains. 

Thus, the inferred isotope ratios provide strong evidence that HD\,101584 was a low-mass, M-type AGB star.

\subsection{The extreme velocity spots}
\label{s:hotspots}

The high-velocity outflow is marked with a number of emission spots along its extent, particularly prominent in the SiO(\mbox{5--4}) data (Fig.~\ref{f:maps}). The end-points at about 4\arcsec\,W and 4\arcsec\,E of the central star are visible as distinct features at the extreme velocities $\upsilon_{\rm LSR}$\,$\approx$\,--100 and 185\,km\,s$^{-1}$, respectively, in many of the detected molecular lines (we refer to \citet{olofnyma99}). Notably, the H$_2$CO (3$_{03}$--2$_{02}$, 3$_{22}$--2$_{21}$, and 3$_{21}$--2$_{20}$) line emissions emanate most strongly from these extreme-velocity spots (EVSs) (Fig.~\ref{f:maps}). This suggests special conditions, either in chemical or excitation terms, and the presence of shocked gas is likely. With an inclination angle of 10$^\circ$, the physical distances between the star and the EVSs are $\approx$\,3$\times$10$^{17}$\,cm, that is, $\approx$\,0.1\,pc or 20000\,au.

\subsection{The detection of methanol}

\begin{figure}
   \centering
  \includegraphics[width=8cm]{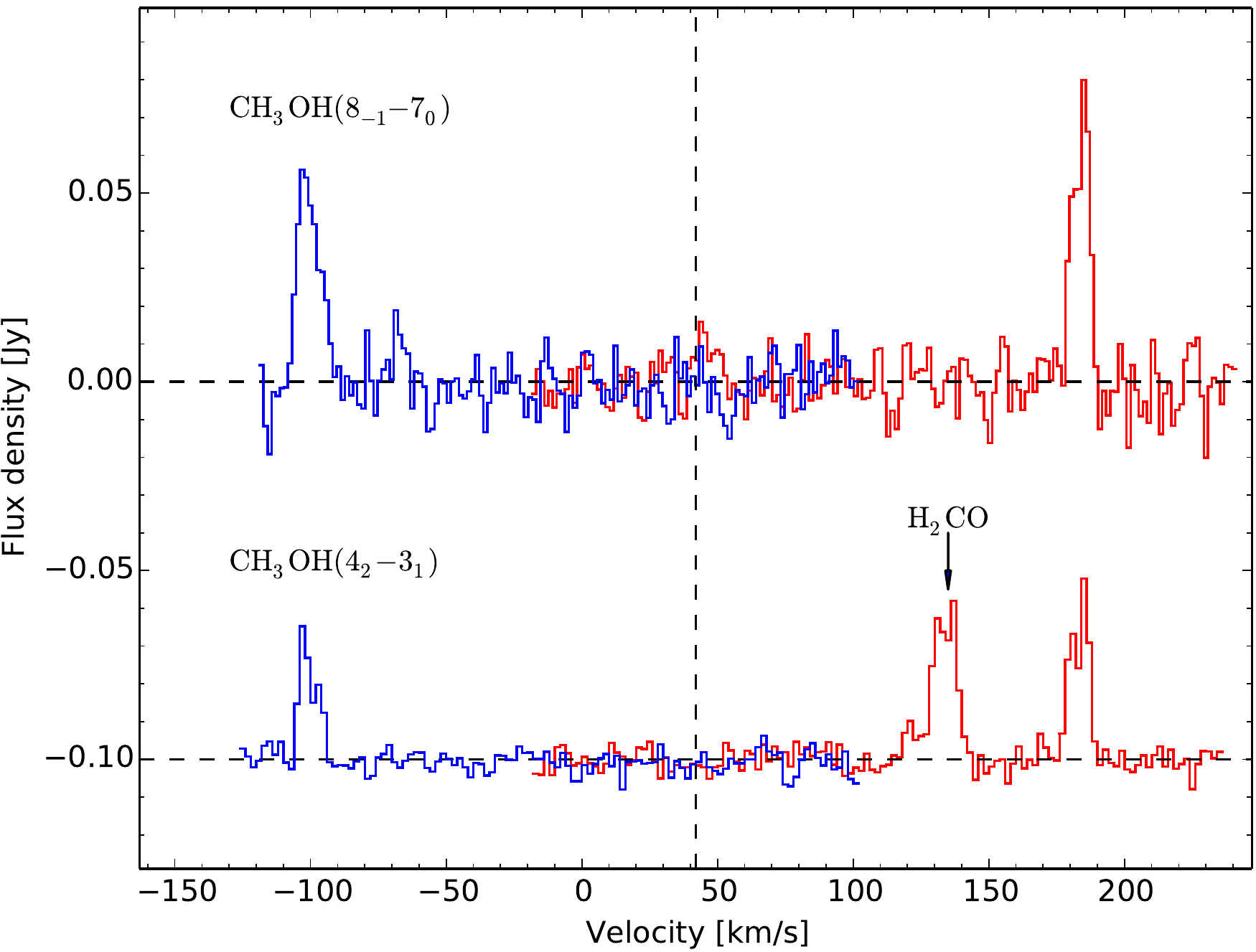} 
      \caption{Methanol lines detected towards the 4\arcsec\ W (blue) and 4\arcsec\ E (red) EVSs; 4$_{2}$--3$_{1}$ (bottom), 8$_{-1}$--7$_{0}$ (top). The lines are integrated over 1\arcsec\ $\times$ 1\arcsec\ areas centered on the spots. The feature at $\approx$\,130\,km\,s$^{-1}$ is due to the H$_2$CO(3$_{22}$--2$_{21}$) line. The vertical dashed line marks the estimated systemic velocity. A Local Standard of Rest velocity scale is used.
              }
         \label{f:ch3ohlines}
   \end{figure}

We detect two CH$_3$OH lines, the $J_K$\,=\,4$_2$--3$_1$ line at 218.440\,GHz and the 8$_{-1}$--7$_0$ line at 229.759\,GHz (Fig.~\ref{f:ch3ohlines}). In the frequency range covered by our ALMA data, these are also the strongest lines in, for example, star-forming regions \citep{nummetal98}, where the 8$_{-1}$--7$_0$ line has a tendency to show maser emission \citep{kaleetal02}, but there is no indication of this in our data. At the sensitivity obtained, the CH$_3$OH line emission comes exclusively from the EVSs (Fig.~\ref{f:maps}), further supporting the idea that the conditions are special in these regions. To our knowledge, this is the first detection ever of methanol in an AGB-related object. We note that IRAS\,19312+1950 has been detected in methanol, but it has been convincingly shown to be a young object rather than an AGB star \citep{cordetal16}. Likewise, methanol was detected towards IRC+10420, but this is an evolved massive star \citep{quinetal16b}.

\subsection{Molecular abundances}

We need a physical model for an EVS to derive the molecular abundances through a radiative transfer analysis. The observational data provide limited information here except that the emission is largely confined to a region of size $\la$\,1\arcsec, and the emission from the rarer species seems to come from a region at least two times smaller than this. The geometry is also uncertain. In order to get at least order-of-magnitude estimates we assume a spherical clump with an inner higher-density, higher-temperature region (0$\farcs$15 in diameter) surrounded by an outer lower-density, lower-temperature envelope (1\arcsec\ in diameter; required to mainly reproduce the CO line intensities); we refer to Table~\ref{t:model}. Radiative excitation due to central star light (too distant) and dust emission inside the clump (too low optical depth) can be safely ignored.

Based on this physical model we solve the radiative transfer using an Accelerated-Lambda-Iteration code, taking into account excitation through collisions with H$_2$. Collisional coefficients for CH$_3$OH were taken from \citet{pottetal04}. We have not included the torsionally excited states of methanol, which may lead to an underestimate of its abundance by up to a factor of two. We use as input the line intensities for the 4\arcsec\,E EVS where the emission is slightly stronger (Table~\ref{t:obs}). The size of the higher-density region is constrained by requiring that the $^{28}$SiO/$^{29}$SiO abundance ratio equals 20 (the solar value) since an AGB star is not expected to alter this ratio. Its size fits well the observed sizes of the most intense molecular line emissions of, for example, H$_2$CO and CH$_3$OH. The densities and kinetic temperatures are reasonably constrained by the observed CO, H$_2$CO, and CH$_3$OH line intensities (Table~\ref{t:model}).

The resulting fractional abundances with regards to H$_2$ are listed in Table~\ref{t:model}. The $^{12}$CO abundance (5$\times$10$^{-4}$) is at the level expected in an O-rich circumstellar gas (full association of CO and solar values for O and C results in a fractional CO abundance of 5$\times$10$^{-4}$). Furthermore, the $^{12}$CO/$^{13}$CO ratio is $\approx$\,15, that is, in very good agreement with the value derived in Sect.~\ref{s:nature}. Importantly, this means that the EVS material is dominated by circumstellar gas (possibly swept-up from the previous AGB wind), not by swept-up interstellar material. This is strengthened by the H$_2$CO/H$_2$$^{13}$CO line intensity ratio, 12$^{+12}_{-6}$. The C$^{16}$O/C$^{18}$O ratio of 210 (the solar value is 480, \citet{scotetal06}) is somewhat low since AGB stars are expected to destroy rather than produce $^{18}$O, but considering our simple model we draw no conclusions based on this result.  

The estimated gas mass of the 4\arcsec\,E EVS is 4$\times$10$^{-3}$\,M$_\odot$. A crude estimate shows that such a clump would produce a 1.3\,mm continuum flux density that is lower than the noise limit in our ALMA data. 

\begin{table}
\caption{Radiative transfer results}
\centering
\begin{tabular}{l l l l}
\hline \hline
Inner: & $n_{\rm H_2}$\,=\,5$\times$10$^6$\,cm$^{-3}$, & $T_{\rm k}$\,=\,200\,K & $R$\,=\,0{\farcs}15 \\  
Outer: & $n_{\rm H_2}$\,=\,5$\times$10$^5$\,cm$^{-3}$, & $T_{\rm k}$\,=\,60\,K & $R$\,=\,1\arcsec\ \\  
\hline \hline 
Species        & $f_{\rm X,inner}$  &  $f_{\rm X,outer}$\\
\hline 
$^{12}$CO  & 5.0$\times$10$^{-4}$ & 5.0$\times$10$^{-4}$ \\
$^{13}$CO  & 3.3$\times$10$^{-5}$ & 3.3$\times$10$^{-5}$\\
C$^{18}$O & 2.4$\times$10$^{-6}$  & 2.4$\times$10$^{-6}$\\
$^{13}$CS & 1.0$\times$10$^{-7}$ \\
SO & 1.0$\times$10$^{-6}$  \\
SiO  & 5.0$\times$10$^{-6}$ \\
$p$-H$_2$CO & 6.0$\times$10$^{-7}$  \\
$o$-H$_2^{13}$CO &  1.8$\times$10$^{-7}$ \\
$E$-CH$_3$OH & 1.6$\times$10$^{-6}$  & 2.0$\times$10$^{-7}$\\
  \hline
\end{tabular}
\label{t:model}
\end{table}

\subsection{Chemistry}

We focus here on the results for the inner region of the 4\arcsec\,E EVS. The SO (1$\times$10$^{-6}$), SiO (5$\times$10$^{-6}$), and CS (1$\times$10$^{-6}$; after correcting the $^{13}$CS abundance by the estimated $^{12}$C/$^{13}$C\,$\approx$\,10) abundances all lie in the range reported for AGB CSEs \citep{danietal16, schoetal13, velietal17}, while the H$_2$CO abundance (2$\times$10$^{-6}$; assuming an ortho-to-para ratio of 3) is about an order of magnitude higher than in IK~Tau \citep{velietal17}. CH$_3$OH has an estimated abundance of 3$\times$10$^{-6}$ (assuming an $E$-to-$A$ ratio of 1), almost two orders of magnitude higher than towards the supergiant IRC+10420 \citep{quinetal16b}. However, the large distance between HD\,101584 and the EVS (3$\times$10$^{17}$\,cm) combined with a reasonable expansion velocity of the AGB wind (15\,km\,s$^{-1}$) indicate a time scale of $\approx$\,6500\,yr. This strongly suggests that due to photodissociation all detected species, except CO which is self-shielding, have their origin in the EVS.

A comparison with the results for OH231.8+4.2, a post-AGB object with a rich molecular setup (incl. SO, SiO, CS, and H$_2$CO) and a number of characteristics similar to those of HD\,101584 \citep{velietal15}, shows that also here some of the species are particularly abundant in regions where shocks are likely present, although these are not associated with the outer extremes of its high-velocity outflow. Another interesting comparison can be made with the results for the high-velocity outflows of young stellar objects (YSOs), where in particular the features at the extreme velocities resemble the EVS emission features of HD\,101584 \citep{bachetal91a, bachetal91c}. Interestingly, the detected abundant molecules, including CH$_3$OH, are largely to be the same \citep{tafaetal10}; as are the masses and temperatures of the clumps.

It is therefore tempting to compare with the work on the chemistry of such outflows, for example, that of L1157-B1 \citep{codeetal10, beneetal13}. In this case a chemical model where gas-grain interaction, including freeze-out and chemical processing over $\approx$\,10$^5$\,yr, and subsequent release of the formed species by a C-shock, works reasonably well to explain the observed abundances \citep[e.g.,][]{vitietal11}. However, there are notable differences with HD\,101584.  In the latter, the shock works on pre-existing circumstellar grains, not interstellar grains coated in a proto-stellar environment. Further, in HD\,101584 the observed CH$_3$OH/H$_2$CO abundance ratio is $\approx$\,1 as opposed to a value of $\approx$\,20 for L1157-B1 \citep{beneetal13}. Finally, H$_2$S is detected in L1157-B1 \citep{holdetal16}, while this is not the case for the EVSs of HD\,101584 despite the fact that the H$_2$S(2$_{20}$--2$_{11}$) line, emanating from the central region of HD\,101584, is almost as strong as the $^{12}$CO(2--1) line (about 30\% of its strength) (Olofsson et al., in prep). An alternative explanation could be evaporation of pre-existing planetary system objects, but also here the observed CH$_3$OH/H$_2$CO abundance ratio is very different, for example, $\approx$\,10 in comets \citep{devaetal13}. 
Lacking an obvious explanation, but building on the similarity with the YSO high-velocity outflows, we propose that the circumstellar grains around HD\,101584 have had time to develop some surface chemistry, the result of which is liberated when the high-velocity gas hits the circumstellar medium.

\section{Conclusions}

We have, for the first time, detected CH$_3$OH towards an AGB-related object, HD\,101584. Among other things, this is interesting since the detection of methanol is normally taken as being characteristic of star-forming activity. The detections of CS, SO, SiO, H$_2$CO, and CH$_3$OH in the EVSs of HD\,101584 follow a very similar pattern of molecular detections in the extreme velocity flows of YSOs. However, there are significant differences both in the environmental conditions and in the observed (relative) abundances. Nevertheless, based on the similarity, we propose that the detected molecular species in the EVSs have their origin in a post-shock chemistry where circumstellar grains play a role; the details of this situation, however, remain to be elucidated. It would be interesting to perform searches for CH$_3$OH in other post-AGB objects where shocks are likely to be present, for example, the water fountain sources \citep[e.g.,][]{tafoetal14}. Finally, we have added evidence that HD\,101584 is a post-AGB object, the remnant of a solar-mass M-type AGB star.

\begin{acknowledgements}
HO and WV acknowledge support from the Swedish Research Council. WV acknowledges support from the ERC through consolidator grant 614264.
This Letter makes use of the following ALMA data: ADS/JAO.ALMA\#2012.1.00248.S and \#2015.1.00078.S. ALMA is a partnership of ESO (representing its member states), NSF (USA) and NINS (Japan), together with NRC (Canada) and NSC and ASIAA (Taiwan), in cooperation with the Republic of Chile. The Joint ALMA Observatory is operated by ESO, AUI/NRAO and NAOJ.
This paper makes use of the following APEX data: O-093.F-9307 and O-096.F-9303.The Atacama Pathfinder EXperiment (APEX) is a collaboration between the Max-Planck-Institut f{\"u}r Radioastronomie, the European Southern Observatory, and the Onsala Space Observatory.
\end{acknowledgements}



\newpage

\begin{appendix}

\section{Observational results}

\begin{table*}
\caption{Observational results towards the center of HD\,101584}
\centering
\begin{tabular}{l c c c c c}
    \hline \hline
    Line        & Tel. & $S$  & $\int S {\rm d}\upsilon$ & $\Delta \upsilon$ & $\upsilon_{\rm c}$ \\
                &      & [Jy] & [Jy\,km\,s$^{-1}$]       & [km\,s$^{-1}$]    &  [km\,s$^{-1}$]\\
   \hline
C$^{17}$O(2--1)\,$^1$  & APEX      &  0.28\,(0.06)\phantom{0}\phantom{0}   &  1.8\,(0.6)\phantom{0}\phantom{0} & 6.4\,(1.7) & 41.3\,(3.5)\\
C$^{18}$O(2--1)\,$^1$ & APEX       &  1.3\,(0.09)\phantom{0}\phantom{0}\phantom{0}   & 9.2\,(1.0)\phantom{0}\phantom{0} & 7.1\,(0.6) & 41.5\,(2.4)\\
$^{13}$C$^{17}$O(2--1) &ALMA &  0.036\,(0.005)  &  0.14\,(0.03) &  3.9\,(0.7) & 41.7\,(0.3)\\
  \hline
\end{tabular}
\label{t:isodata}
\tablefoot{ The values within parentheses give the uncertainties in the Gaussian fits used to determine the line parameters. (1) A Gaussian line decomposition has been used to identify the narrow feature from the central region. }
\end{table*}

\begin{table*}
\caption{Observational results at the 4\arcsec\ E EVS. The ALMA data are integrated over an 1\arcsec\ $\times$ 1\arcsec\ area centered on the spot.}
\centering
\begin{tabular}{l c c c }
    \hline \hline
    Line        & Telescope & $S\,^1$  &  $\Delta \upsilon\,^1$ \\
                &      & [Jy] &  [km\,s$^{-1}$] \\
   \hline
$^{12}$CO(1--0)  & SEST & 0.81\phantom{0} \\
$^{12}$CO(2--1)  & ALMA & 3.0\phantom{0}\phantom{0}   & 15.7\\
$^{12}$CO(3--2)\,$^2$  & APEX & 3.4\phantom{1}\phantom{0}\\
$^{12}$CO(4--3)\,$^3$  & APEX & 3.0\phantom{1}\phantom{0}\\
$^{13}$CO(1--0)  & SEST & 0.16\phantom{0}\\
$^{13}$CO(2--1)  & ALMA & 1.1\phantom{0}\phantom{0}     & \phantom{1}10.0\\
C$^{18}$O(2--1)  & ALMA  & 0.061 & \phantom{1}8.7\\
$^{13}$CS(5--4)  & ALMA  & 0.047 & \phantom{1}7.3  \\
SO(5$_5$--4$_4$) & ALMA & 0.074 & \phantom{1}9.4 \\
SO(6$_5$--5$_4$) & ALMA & 0.093  & \phantom{1}8.9\\
SiO(5--4)        & ALMA & 0.50\phantom{0}   & \phantom{1}7.8\\
$^{29}$SiO(5--4) & ALMA & 0.19\phantom{0}       & \phantom{1}6.8\\
H$_2$CO(3$_{03}$--2$_{02}$) & ALMA & 0.14\phantom{0} & \phantom{1}7.2 \\
H$_2$CO(3$_{22}$--2$_{21}$) & ALMA & 0.042 & \phantom{1}7.6 \\
H$_2$CO(3$_{21}$--2$_{20}$) & ALMA & 0.041 & \phantom{1}6.5 \\
H$_2^{13}$CO(3$_{12}-2_{11}$) & ALMA  & 0.03\phantom{0} & \phantom{1}8.3 \\
CH$_3$OH(4$_2$--3$_1$) & ALMA & 0.039 & \phantom{1}6.5 \\
CH$_3$OH(8$_{-1}$--7$_0$) & ALMA & 0.065 & \phantom{1}7.4  \\
  \hline
\end{tabular}
\label{t:obs}
\tablefoot{(1) The 1$\sigma$ noise levels lie in the range 2-4\,mJy, therefore the uncertainties in the line parameters are dominated by uncertainties in the cleaning process and the somewhat complicated brightness distributions. We estimate the uncertainties to be about 25\,\% in the flux density and 1\,km\,s$^{-1}$ in the line width. (2) Corrected for the response of a 18\arcsec\ primary beam. (3) Corrected for the response of a 14\arcsec\ primary beam}
\end{table*}

\end{appendix}

\end{document}